\documentclass[11pt,conference]{IEEEtran}
\usepackage[utf8]{inputenc}
\usepackage[T1]{fontenc}
\usepackage{graphicx,color,mathenv}
\usepackage[cmex10]{amsmath}
\usepackage{cite,amsthm,amssymb}
\usepackage{balance}
\usepackage[ruled,vlined]{algorithm2e}
\usepackage{subfigure,amsfonts}
\usepackage{float}
\usepackage{url}
\usepackage{xcolor}

\usepackage{bbm}
\usepackage{cancel}

\DeclareMathAlphabet\mathbfcal{OMS}{cmsy}{b}{n}

\begin{document}

\title{On the Noise Robustness of Affine Frequency Division Multiplexing: Analysis and Applications}

\author{\IEEEauthorblockN{
Vincent Savaux\IEEEauthorrefmark{1},
Steve Sawadogo\IEEEauthorrefmark{1}\IEEEauthorrefmark{2}, 
Hyeon Seok Rou\IEEEauthorrefmark{3}, 
Giuseppe Thadeu Freitas de Abreu\IEEEauthorrefmark{3}
}
\IEEEauthorblockA{\IEEEauthorrefmark{1}b$<>$com, 1219 Av. des Champs Blancs, 35510 Cesson-S\'{e}vign\'{e}, France}
\IEEEauthorblockA{\IEEEauthorrefmark{2}INSA Rennes, CNRS, IETR - UMR 6164, F-35000, Rennes, France}
\IEEEauthorblockA{\IEEEauthorrefmark{3}School of Computer Sci. and Eng., Constructor University Bremen, Campus Ring 1, 28759 Bremen, Germany}
}


\maketitle

\begin{abstract}
This paper investigates the robustness of affine frequency division multiplexing (AFDM) and orthogonal time frequency space (OTFS) modulation schemes against non-white Gaussian noise, which can model various sources of additive disturbances to the received signal. The proposed approach demonstrates that the performance of these waveforms depends on the ability of the demodulation matrix to whiten the noise—a property that is, in turn, related to the sparsity of the matrix. AFDM is shown to outperform OTFS and orthogonal frequency division multiplexing (OFDM), as its demodulation matrix is generally less sparse than those of the other waveforms. Based on this analysis, several application examples and use cases are presented, such as the use of AFDM and OTFS in narrowband signals or in coexistence with OFDM signals. Finally, simulation results confirm that AFDM achieves better performance than OTFS and OFDM in the presence of non-white noise, with gains exceeding 1 dB in most application scenarios.  

\end{abstract}

\begin{IEEEkeywords}
5G, 6G, coexistence, OFDM, AFDM, OTFS, noise
\end{IEEEkeywords}


\section{Introduction}
\label{sec:intro}

Affine frequency division multiplexing (AFDM) and orthogonal time frequency space (OTFS) modulation schemes have emerged as promising alternatives to orthogonal frequency division multiplexing (OFDM) for sixth-generation (6G) communications, as they better match the technical requirements of future use cases compared to OFDM \cite{tataria22,chafii23}. In particular, they are more robust against highly time- and frequency-selective propagation channels, as they can achieve full diversity \cite{bemani21} in such environments, unlike OFDM. In fact, in OTFS, the transmitted data is mapped into the delay-Doppler domain. The signal is then modulated using an inverse symplectic Fourier transform, followed by a Heisenberg transform \cite{hadani17}. AFDM is based on the discrete affine Fourier transform (DAFT) \cite{erseghe05} rather than the discrete Fourier transform (DFT) used in OFDM \cite{bemani21}. As a result, data symbols are carried by chirps instead of sinusoids, which allows the errors introduced by doubly selective channels to be spread across the time-frequency grid. 

The robustness of OFDM, AFDM, and OTFS against time and frequency selective channels has recently been extensively studied \cite{hashimoto21,das21,huang23,yin22mimo,yin24,cao24arxiv,zhou24arxiv,benzine24arxiv,zheng24,bemani22icassp}, and their performance has been compared in the context of integrated sensing and communications \cite{rou24}. However, to the best of the authors' knowledge, the robustness of these modulation schemes has not been analyzed in the context of additive non-white Gaussian noise. Yet such a general noise model can represent a wide range of disturbances such as filtered white noise at the receiver (including channel equalization), aliasing, interference, jamming, and so on. 

In this paper, the performance of AFDM and OTFS modulation schemes in non-white Gaussian noise environment is investigated. First, we express the (de)modulation process of AFDM and OTFS as precoded OFDM. Then, we analyze the capability of the demodulation matrices of the waveforms to whiten the received noise. We show that it can be directly related to the sparsity of such matrices. In fact, using the central limit theorem, we can show that the lowest the sparsity of the demodulation matrix, the better the capability of the waveform to whiten the noise. In this respect, AFDM offers the best capability regardless of the subcarriers number, while OTFS is limited due to its inherent dependence  
on the delay-Doppler grid size.  

Based on this analysis, we provide examples of applications and uses cases where AFDM should outperform OTFS. Due its better noise whitening capability, AFDM is inherently more robust than OTFS and OFDM against threats such as noise injection or jamming from a possible malicious user. Then, it offers better physical layer security. Since this capability is independent of the subcarriers number, AFDM could be used as a resilient modulation for narrowband communication in Internet of things (IoT) or machine-type applications. Otherwise, AFDM could be used in the context of coexistence with OFDM where both waveforms share the same spectrum resource. We introduce what we call multi-waveform frequency division multiple access (FDMA) in which multiple users using different waveforms are allocated to small parts of the available bandwidth. 

The suggested analysis is supported by simulation results showing that AFDM actually outperforms OTFS and OFDM in whitening non-white noise. The simulation parameters are chosen to match the aforementioned applications and use cases. In particular, we show that the bit error rate (BER) performance of AFDM is independent of the chirp parameter, while the results of OTFS can degrade depending on the size of the delay-Doppler grid, especially if the subcarriers number is small. In a doubly dispersive channel with impulse noise, AFDM outperforms OTFS by more than 1~dB in the high signal-to-noise ratio (SNR) range.

The rest of the paper is organized as follows: Section \ref{sec:system_model} introduces the frequency domain (de)modulation expressions for the OTFS and AFDM signals. Then, Section \ref{sec:perf_analysis} is dedicated to the performance analysis of OTFS and AFDM modulation schemes against non-white noise. Examples of applications and use cases are provided in Section \ref{sec:appli}. Simulation results are shown in Section \ref{sec:simu} and conclusions are drawn in Section \ref{sec:conclusion}. 

\textit{Notations}: Boldface $\textbf{a}$ and $\textbf{A}$ is used for vectors and matrices, respectively, while normal font $a$ indicates scalar variables. $\textbf{A}^{*}$, $\textbf{A}^{T}$, and $\textbf{A}^H$ represent the conjugate, the transpose, and the conjugate transpose (Hermitian) of matrix $\textbf{A}$, respectively. The convolution operation is expressed by $(.*.)$, and $. \otimes .$ represents the Kronecker product. The mathematical expectation is denoted by $\mathbb{E}\{.\}$. 


\section{System Model}
\label{sec:system_model} 

This section introduces the input-output relations of the OFDM, OTFS, and AFDM waveforms. Since all are multicarrier signals, a common framework can be considered using a general modulation matrix denoted by $\textbf{Q}$. Note that a more general framework can also be found in \cite{boudjelal25}. Thus, let us consider a multicarrier system consisting of $N$ subcarriers. In OFDM, the vector of transmitted signal is simply obtained by the IDFT of the vector $\textbf{c} \in \mathbb{C}^N$ that contains the data taken from a constellation, leading to: 

\begin{equation}
    \textbf{x}^{\text{of}} = \mathbfcal{F}_{N}^H \textbf{c}, 
    \label{eq:xofdm}
\end{equation} 
where the superscript \emph{of} indicates "OFDM". Moreover, the matrix $\mathbfcal{F}_{N}^H$ is the IDFT matrix of size $N\times N$ containing the element $\frac{e^{2j\pi \frac{mn}{N}}}{\sqrt{N}}$ at entry $(n,m)$. In OTFS, the data is first multiplexed in the delay Doppler grid of size $K \times L$, such that the transmitted signal $\textbf{x}^{\text{ot}}$ is obtained as: 

\begin{align}
    \textbf{x}^{\text{ot}} =& \text{vec}\left( \mathbfcal{F}_{K}^H (\mathbfcal{F}_{K} \textbf{C} \mathbfcal{F}_{L}^H) \right) \nonumber \\ 
    =& (\mathbfcal{F}_{L}^H \otimes \textbf{I}_K) \textbf{c}, 
    \label{eq:xotfs}
\end{align}
where $\textbf{C}$ is the $K \times L$ matrix containing the data symbols such that $\textbf{c} = \text{vec}(\textbf{C})$, and $N = LK$. Interestingly, note that $L=N$ corresponds to OFDM, whereas $L=1$ corresponds to single carrier-FDM. Otherwise, the value of $L$ allows for a trade-off between accuracy in the delay (\emph{i.e.} $K$) and Doppler domains \cite{rou24}. 
In AFDM, the subcarriers are chirps instead of sinusoids as in OFDM. However, the transmitted AFDM signal $\textbf{x}^{\text{af}}$ can be easily represented in a matrix form as follows: 

\begin{equation}
    \textbf{x}^{\text{af}} = \boldsymbol{\Lambda}_q \mathbfcal{F}_{N}^H \boldsymbol{\Lambda}_\alpha \textbf{c}, 
    \label{eq:xafdm}
\end{equation}
where $\boldsymbol{\Lambda}_q$ and $\boldsymbol{\Lambda}_\alpha$ are both $N \times N$ diagonal matrices containing the elements $e^{j\frac{\pi q n^2}{N}}$ and $e^{j\frac{\pi \alpha n^2}{N}}$, with $n=0,1,..,N-1$. The parameters $q,\alpha \in \mathbb{R}$ can be set to achieve full diversity in time and frequency selective channels \cite{bemani21} (note that in some papers, $q$ and $\alpha$ correspond to $c_1$ and $c_2$ where $q=2Nc_1$ and $\alpha = 2Nc_2$). It can be noted that in the case where $|q|=1$, the AFDM corresponds to the OCDM \cite{ouyang16}, and $q=0$ leads to the OFDM modulation scheme. 
More generally, it is worth mentioning that the transmitted signal $\textbf{x}$ in (\ref{eq:xofdm})-(\ref{eq:xafdm}) can be rewritten as a precoded OFDM as: 

\begin{equation}
    \textbf{x} = \mathbfcal{F}_{N}^H \textbf{Q} \textbf{c}, 
    \label{eq:xgen}
\end{equation}
where $\textbf{Q} = \textbf{I}_N$ in OFDM, $\textbf{Q} = \mathbfcal{F}_{N}(\mathbfcal{F}_{L}^H \otimes \textbf{I}_K)$ in OTFS, and $\textbf{Q} = \mathbfcal{F}_{N}\boldsymbol{\Lambda}_q \mathbfcal{F}_{N}^H \boldsymbol{\Lambda}_\alpha$ in AFDM, can be interpreted as a precoding matrix. 
In the following, we denote by $\textbf{z} = \textbf{Q} \textbf{c}$ the precoded data. 
Then, for any modulation scheme, the general input-output relation in SISO systems can be expressed as

\begin{equation}
    \textbf{y} = \underbrace{\sum_{l=0}^{L_c-1}h_l \boldsymbol{\Delta}_{\theta_l}\boldsymbol{\Pi}^l \textbf{x}}_{\textbf{H} \textbf{x}}  + \textbf{w}, 
    \label{eq:receivedsignaly}
\end{equation}
where $\textbf{y}$ is the $N \times 1$ vector of the received signal, and $\textbf{w}$ is the vector of additive noise, which is assumed to be Gaussian but not necessarily white. Such a general non-white noise model can correspond to any disturbance as filtered white noise at the receiver, aliasing, interference, or jamming. Then, $\textbf{H}$ is the $N \times N$ channel matrix, where  $h_l$ is the $l$th channel path coefficient (some of the $h_l$ can be possibly zero if the channel is sparse), and $L_c$ is the channel length. Moreover, $\boldsymbol{\Delta}_{\theta_l}$ is the $N \times N$ diagonal matrix containing the samples $e^{2j\pi \frac{\theta_l n}{N}}$, where $\theta_l$ is the normalized Doppler shift, and $\boldsymbol{\Pi}$ is the forward cyclic-shift matrix. 

The pilot design, channel estimation, and equalization in OFDM \cite{ozdemir07,liu14,savaux17IET}, OTFS \cite{hashimoto21,das21,huang23}, and AFDM \cite{yin22mimo,yin24,cao24arxiv,zhou24arxiv,benzine24arxiv,zheng24,bemani22icassp} has been largely studied in the literature, and not further dealt with in this paper. 
In any case, to recover the data vector $\hat{\textbf{c}}$ from the observation $\textbf{y}$, the usual receiver can be expressed as: 

\begin{equation}
    \hat{\textbf{c}} = \textbf{G}\textbf{y}, 
    \label{eq:hatc}
\end{equation}
where $\textbf{G}$ is the equalization matrix of size $N \times N$. Assuming that the channel can be perfectly equalized (\emph{e.g.} $\textbf{G} = (\textbf{H}^H\textbf{H})^{-1}\textbf{H}^H$ in case of zero-forcing equalization), and using straightforward matrix manipulation, (\ref{eq:hatc}) can be rewritten as



\begin{align}
    \hat{\textbf{c}} =& \textbf{Q}^{-1} \textbf{G}_f \mathbfcal{F}_{N}\textbf{y} \nonumber \\
    =&  \textbf{c} + \textbf{Q}^{-1} \textbf{G}_f \textbf{w}_f, 
    \label{eq:hatcfreq}
\end{align}
where $\textbf{G}_f = \mathbfcal{F}_{N} \textbf{G} \mathbfcal{F}_{N}^H$ and $\textbf{w}_f = \mathbfcal{F}_{N}\textbf{w}$ are the Fourier transforms of the equalization matrix and the noise vector, respectively. From (\ref{eq:hatcfreq}), we notice that, in general, the demodulated data $\hat{\textbf{c}}$ can be expressed as a noisy version of the data $\textbf{c}$. From this observation, we evaluate and compare the performance of the OTFS and AFDM waveforms receivers in next Section.  



\section{Performance of OTFS and AFDM Against Non-White Noise}
\label{sec:perf_analysis} 

 Since non-white Gaussian noise reduces the performance of signal receivers compared to the usual AWGN, the performance (in term of BER) of the waveforms can be directly linked to the capability of the corresponding receivers to whiten the noise. Thus, we evaluate the performance of the different waveforms against non-white noise through the capability of the demodulation matrices $\textbf{Q}^{-1}$ to whiten the reception noise $\textbf{G}_f \textbf{w}_f$. First, we hereby formalize the non-white noise model that is used afterward. 

\subsection{Non-White Noise Model}

Although $\textbf{G}_f \textbf{w}_f$ in (\ref{eq:hatcfreq}) is assumed to be non-white, it is actually Gaussian, so we can rewrite (\ref{eq:hatcfreq}) to highlight the second order moment of $\textbf{G}_f \textbf{w}_f$ as follows: 

\begin{align}
    \hat{\textbf{c}} =& \textbf{c} + \textbf{Q}^{-1} \textbf{G}_f \textbf{w}_f \nonumber \\ 
    =& \textbf{c} + \textbf{Q}^{-1} \boldsymbol{\Gamma}_f^{1/2} \textbf{w}_w, 
    \label{eq:reccolourednoise}
\end{align}
where $\textbf{w}_w$ is a vector of size $N \times 1$ that contains independent and identically distributed (iid) samples of white Gaussian noise with variance $\sigma_w^2$. Moreover, $\boldsymbol{\Gamma}_f$ is the covariance matrix of the non-white noise, such that 

\begin{equation}
    \mathbb{E}\{\textbf{G}_f \textbf{w}_f(\textbf{G}_f \textbf{w}_f)^H\} = \sigma_w^2 \boldsymbol{\Gamma}_f. 
    \label{eq:varnoise}
\end{equation}
To keep the noise model simple and make the mathematical developments tractable, we assume that $\boldsymbol{\Gamma}_f$ is diagonal, with $(\boldsymbol{\Gamma}_f)_{n,n} = \gamma_n^2 \in \mathbb{R}_+$, $n=0,1,..,N-1$, which highlights the fact that the resulting noise $\boldsymbol{\Gamma}_f^{1/2} \textbf{w}_w$ is non-white. It may be an oversimplified model to capture all kinds of disturbance, in particular if the channel is time selective (in that case $\mathbfcal{F}_{N}\textbf{H}\mathbfcal{F}_{N}^H$ and in turn $\textbf{G}_f$ and $\boldsymbol{\Gamma}_f^{1/2}$ are not diagonal), but it covers numerous cases as frequency selective channel and low Doppler (\emph{e.g.} quasi-static channels), filtering of white noise at the receiver, aliasing, interference, or jamming. The relevance of such an approximation is further evaluated through simulations in Section \ref{sec:simu}. 

It should be noted that the OFDM demodulation matrix $\textbf{Q} = \textbf{I}_N$ is unable to whiten the non-white noise vector $\textbf{G}_f \textbf{w}_f$. In the following, we investigate the capability of the OTFS and AFDM demodulation to whiten the noise. To this end, we link this capability (and then the waveform performance) to the sparsity of the demodulation matrix $\textbf{Q}$. In fact, according to the central limit theorem applied to the rows of $\textbf{Q}^{-1} \boldsymbol{\Gamma}_f^{1/2} \textbf{w}_w$, the lower the sparsity of $\textbf{Q}^{-1}$, the better the capability to whiten the noise.     



\subsection{OTFS}

In the case of OTFS demodulation, the $(u,v)$th entry of $\textbf{Q}^{-1}$ is given by: 

\begin{align}
    (\textbf{Q}^{-1})_{u,v} =& ((\mathbfcal{F}_{L} \otimes \textbf{I}_K) \mathbfcal{F}_{N}^H)_{u,v} \nonumber \\
    =& \sum_{p=0}^{N-1}(\mathbfcal{F}_{L} \otimes \textbf{I}_K)_{u,p}(\mathbfcal{F}_{N}^H)_{p,v}, 
    \label{eq:QOTFS1}
\end{align}
where the elements $(\mathbfcal{F}_{L} \otimes \textbf{I}_K)_{u,p}$ can be expressed as 

\begin{equation}
    (\mathbfcal{F}_{L} \otimes \textbf{I}_K)_{u,p} = \begin{cases}
    \omega_{L}^{\mu \nu}, & \text{ if } (u-p) \text{ mod } K = 0 \\
    0, & \text{ else }
    \end{cases}, 
    \label{eq:FLIK}
\end{equation}
with $\mu = \lfloor u/K \rfloor$ and $\nu = \lfloor p/K \rfloor$. Then, by setting $r = u \text{ mod } K$ and $p = r + Kp'$ where $p' = 0,1,..,L-1$, (\ref{eq:QOTFS1}) can be developed using (\ref{eq:FLIK}) as 

\begin{align}
    (\textbf{Q}^{-1})_{u,v} =& \sum_{p'=0}^{L-1}(\mathbfcal{F}_{L} \otimes \textbf{I}_K)_{u,r+Kp'}(\mathbfcal{F}_{N}^H)_{r+Kp',v} \nonumber \\
    =& \sum_{p'=0}^{L-1}\omega_{L}^{\mu p'}\omega_{N}^{-v(r+Kp')}\nonumber \\
    =& \frac{1}{\sqrt{LN}}\sum_{p'=0}^{L-1} e^{-2j\pi \frac{\mu p'}{L}}e^{2j\pi \frac{v (r+Kp')}{N}}. 
    \label{eq:QOTFS2}
\end{align}
After straightforward manipulations, (\ref{eq:QOTFS2}) simplifies to: 

\begin{align}
    (\textbf{Q}^{-1})_{u,v} =& \frac{1}{\sqrt{LN}}e^{2j\pi \frac{vr}{N}} \sum_{p'=0}^{L-1} e^{2j\pi \frac{p'(v-\mu)}{L}} \nonumber\\
    =& \begin{cases} 
    \frac{L}{\sqrt{LN}}e^{2j\pi \frac{vr}{N}}, & \text{ if }  (v-\mu) \text{ mod } L = 0 \\
    0, & \text{ else }
    \end{cases}. 
    \label{eq:QOTFS3}
\end{align}

Ultimately, the $u$th element of the OTFS-based demodulated non-white noise $(\textbf{Q}^{-1}\boldsymbol{\Gamma}_f^{1/2} \textbf{w}_w)_u$ is given by 

\begin{align}
    (\textbf{Q}^{-1}\boldsymbol{\Gamma}_f^{1/2} \textbf{w}_w)_u =& \sum_{v=0}^{N-1}(\textbf{Q}^{-1})_{u,v} (\boldsymbol{\Gamma}_f \textbf{w}_f)_v \nonumber \\
    =& \sum_{v=0}^{N-1}\frac{L}{\sqrt{LN}}e^{2j\pi \frac{vr}{N}} (\boldsymbol{\Gamma}_f^{1/2} \textbf{w}_w)_v \nonumber \\
    =& \sum_{p''=0}^{K-1}\frac{L}{\sqrt{LN}}e^{2j\pi \frac{r}{K}(r'+Lp')} (\boldsymbol{\Gamma}_f^{1/2} \textbf{w}_w)_{r'+Lp'} 
    \label{eq:otfsnoise}
\end{align}
where $v = r'+Lp''$ and $r' = \lfloor \frac{u}{L} \rfloor$. Since the elements of $\textbf{w}_w$ are iid, in turn those of $\boldsymbol{\Gamma}_f^{1/2} \textbf{w}_w$ are independent but with different variances. Then the capability of $\textbf{Q}^{-1}$ to whiten the noise applies for large $K$ value, \emph{i.e.} when the central limit theorem holds. Thus, it can be noticed that the larger $L$, the larger the sparsity of the matrix $\textbf{Q}^{-1}$. Thus, increasing $L$ improves the robustness of OTFS against Doppler, but at the same time reduces the capability of $\textbf{Q}^{-1}$ to whiten the noise. 



\subsection{AFDM}

In the case of AFDM demodulation, the $(u,v)$th entry of $\textbf{Q}^{-1}$ is given by: 

\begin{align}
    (\textbf{Q}^{-1})_{u,v} =& (\boldsymbol{\Lambda}_\alpha^H\mathbfcal{F}_{N}\boldsymbol{\Lambda}_q^H \mathbfcal{F}_{N}^H )_{u,v} \nonumber \\
    =& (\boldsymbol{\Lambda}_\alpha^H)_{u,u}(\mathbfcal{F}_{N}\boldsymbol{\Lambda}_q^H \mathbfcal{F}_{N}^H )_{u,v}, 
    \label{eq:QAFDM1}
\end{align}
since $\boldsymbol{\Lambda}_\alpha^H$ is diagonal. Thus, evaluating $\textbf{Q}^{-1}$ and its sparsity is equivalent to analyzing $\mathbfcal{F}_{N}\boldsymbol{\Lambda}_q^H \mathbfcal{F}_{N}^H$ in (\ref{eq:QAFDM1}). Because $\boldsymbol{\Lambda}_q^H$ is diagonal, $\mathbfcal{F}_{N}\boldsymbol{\Lambda}_q^H \mathbfcal{F}_{N}^H$ is circulant by Fourier transform, and hence totally defined by its first column. Then, the $u$th element of the vector corresponding to the first column of $\mathbfcal{F}_{N}\boldsymbol{\Lambda}_q^H \mathbfcal{F}_{N}^H$ is obtained by the DFT of the elements of $\boldsymbol{\Lambda}_q^H$ as follows:  

\begin{equation}
(\mathbfcal{F}_{N}\boldsymbol{\Lambda}_q^H \mathbfcal{F}_{N}^H)_{u,0} = \frac{1}{\sqrt{N}} \sum_{k=0}^{N-1} e^{-j\pi \frac{qk^2}{N}} e^{-2j\pi \frac{ku}{N}}. 
\label{eq:QAFDM2}
\end{equation}
In a general case where $q\in \mathbb{R}$, the generalized quadratic Gauss sum in (\ref{eq:QAFDM2}) cannot be simplified, except if $q$ and $N/q$ are integers. In that case, it has been shown in \cite{savaux24ieeetcom} that $(\mathbfcal{F}_{N}\boldsymbol{\Lambda}_q^H \mathbfcal{F}_{N}^H)_{u,0}$ is sparse with $N/q$ non-zero entries that are evenly distributed. Otherwise, we can conclude from (\ref{eq:QAFDM2}) that $\mathbfcal{F}_{N}\boldsymbol{\Lambda}_q^H \mathbfcal{F}_{N}^H$ is generally non-sparse, such as shown in Appendix. Hence, the capability of the demodulation matrix $\textbf{Q}^{-1}$ to whiten non-white noise should be greater in AFDM than in OTFS. Furthermore, such a feature can be independent of $N$ in AFDM (except in special cases where $N/q$ is integer), while it is directly related to the size of the grid $K \times L =N$ in OTFS. This result is further shown through simulations in Section \ref{sec:simu}. Prior to that, we introduce some applications and use cases where the ability of AFDM to whiten non-white noise could be relevant.

\section{Applications and Use Cases}
\label{sec:appli}
This section presents applications and use cases based on the proposed analysis of the robustness of the waveforms against non-white noise. 


\subsection{Physical Layer Security}

A direct consequence of the better capability of OTFS and AFDM to whiten noise compared with OFDM is their better inherent robustness against noise injection, jamming, or interference, as soon as the source of disturbance is non-white. In a scenario where a jammer intends to inject, for instance, an impulsive noise in the legitimate communication, OTFS and AFDM offers better physical layer security than OFDM. Moreover, such a capability is related to $K \leq N$ in OTFS (\ref{eq:otfsnoise}), \emph{i.e.} the dimension of the delay-Doppler grid, and directly to $N$ in AFDM (\ref{eq:QAFDM2}), so one can argue that AFDM could better cope with both doubly dispersive channels and threats from a possible jammer than OTFS. 


\subsection{Narrowband Communications} 

Another consequence of the aforementioned remark is that the dimension of the delay-Doppler grid in OTFS is increasingly constrained as $N$ decreases since $N = K \times L$. Thus, both noise whitening and delay-Doppler diversity cannot be guaranteed when $N$ is small. In contrast, the Doppler diversity in AFDM is related to $q$ independently of $N$. Therefore, AFDM should be more adapted than OTFS and OFDM to narrowband signal communications while guaranteeing delay-Doppler diversity. The corresponding use cases could include machine type communication and IoT, typically using NB-IoT standard where the signal is composed of only up to twelve subcarriers.   


\subsection{OFDM-AFDM/OTFS Coexistence: Multi-Waveform FDMA}

In the following, we present two applications in which AFDM (and possibly OTFS) and OFDM can share frequency resources. This can be especially relevant in use cases of coexistence between OFDM and AFDM (or OTFS) where the robustness of some users against channel or interference can be strengthened using post-OFDM signal for dedicated applications.      




\begin{figure}[tbp] 
\centering
  \includegraphics[width=\columnwidth]{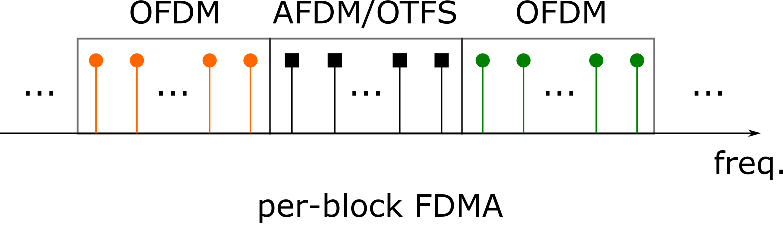} 
\caption{ Multi-waveform FDMA schemes where AFDM/OTFS and OFDM shares adjacent frequency resource, using a per-block mapping. } 
\label{fig:fdma} 
\end{figure} 

In SISO OFDM, multiple users can share the available spectrum using FDMA. We extend this principle by introducing the multi-waveform FDMA, in which sub-signals modulated with different waveforms (\emph{e.g.} OFDM and AFDM) are mapped to contiguous (possibly narrowband) blocks of subcarriers in the frequency domain. Using the orthogonality of subcarriers of any modulation scheme in (\ref{eq:xgen}), the transmitted signal can be expressed as:    


\begin{equation}
    \textbf{x} = \mathbfcal{F}_{N}^H \underbrace{[\textbf{z}_0^T,\textbf{z}_1^T, ..]^T}_{\text{size} N} , 
    \label{eq:xmultiuesiso}
\end{equation}
where $\textbf{z}_0,\textbf{z}_1,..$ represent sub-blocks of signals in the frequency domain (reminding that $\textbf{z} = \textbf{Q} \textbf{c}$ is the precoded data in the frequency domain), each of which can be modulated with OFDM, AFDM, or OTFS. Typically in 3GPP, the size of the blocks can be 12 subcarriers corresponding to the granularity of one resource block. The principle of the multi-waveform FDMA with per-block mapping is illustrated in Fig. \ref{fig:fdma}. According to the possibly small sizes of the blocks, AFDM should be preferred to OTFS to guarantee delay-Doppler diversity and at the same time, robustness against possible jamming. At the receiver side, the different signals can be recovered after a DFT of size $N$, and then a specific demodulation matrix $\textbf{Q}^{-1}$ can be applied to each block according to the considered waveform.

\section{Simulation and Discussion}
\label{sec:simu}


\subsection{Noise Whitening Capability}

\begin{figure}[tbp]
\centering
  \subfigure[impulse noise.]{\includegraphics[width=\columnwidth]{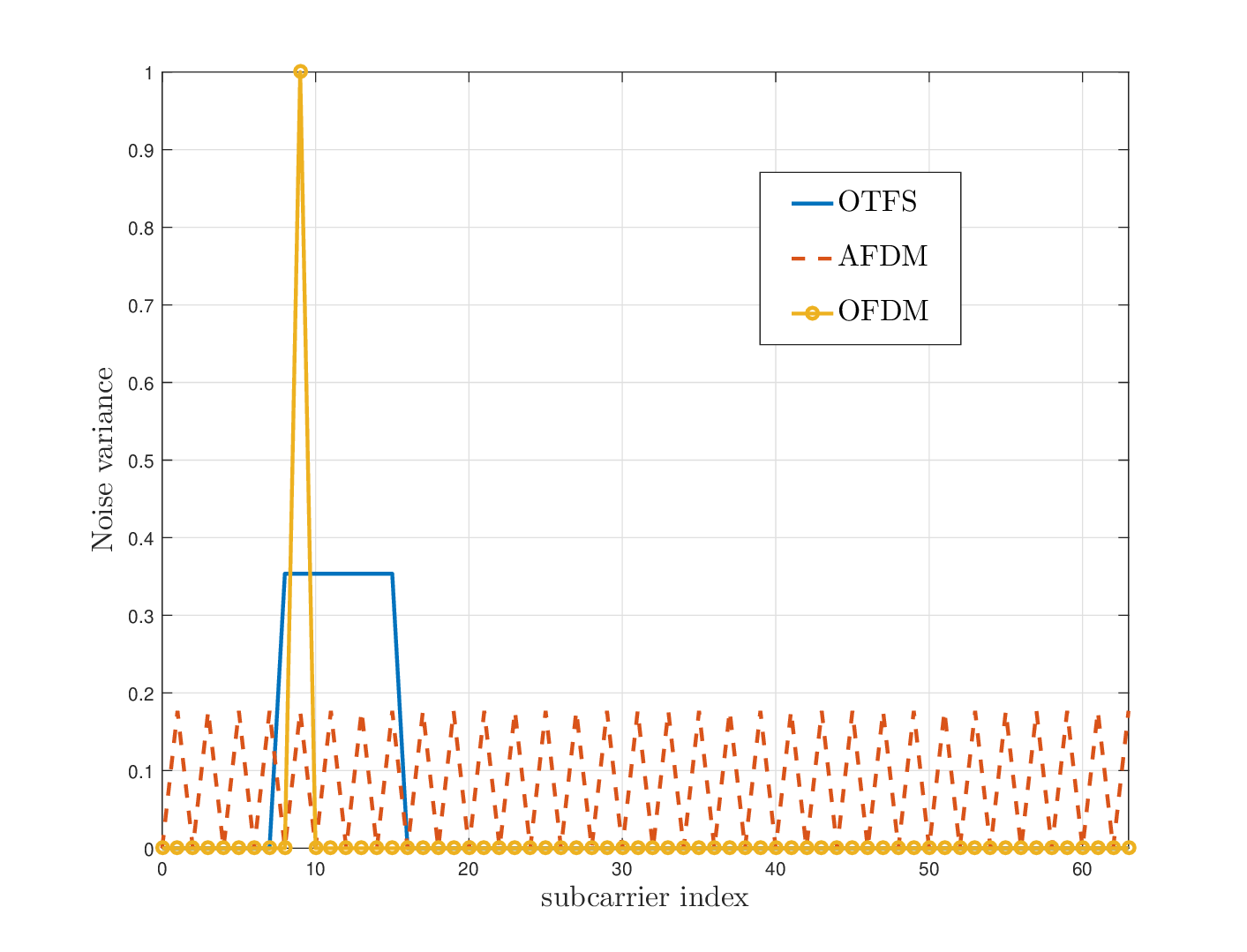}} \quad
  \subfigure[Interferer.]{\includegraphics[width=\columnwidth]{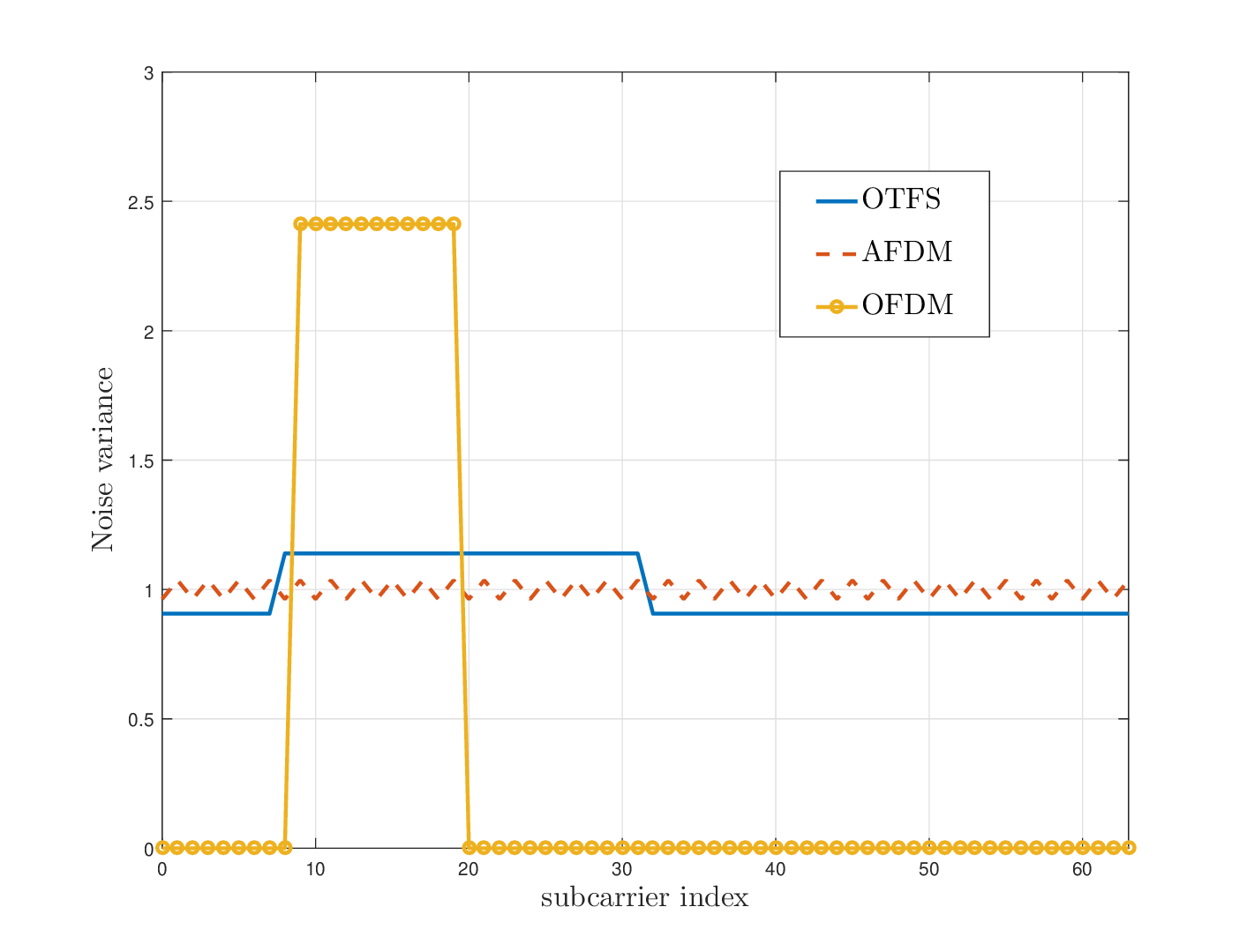}} \quad 
  \subfigure[Equalized noise.]{\includegraphics[width=\columnwidth]{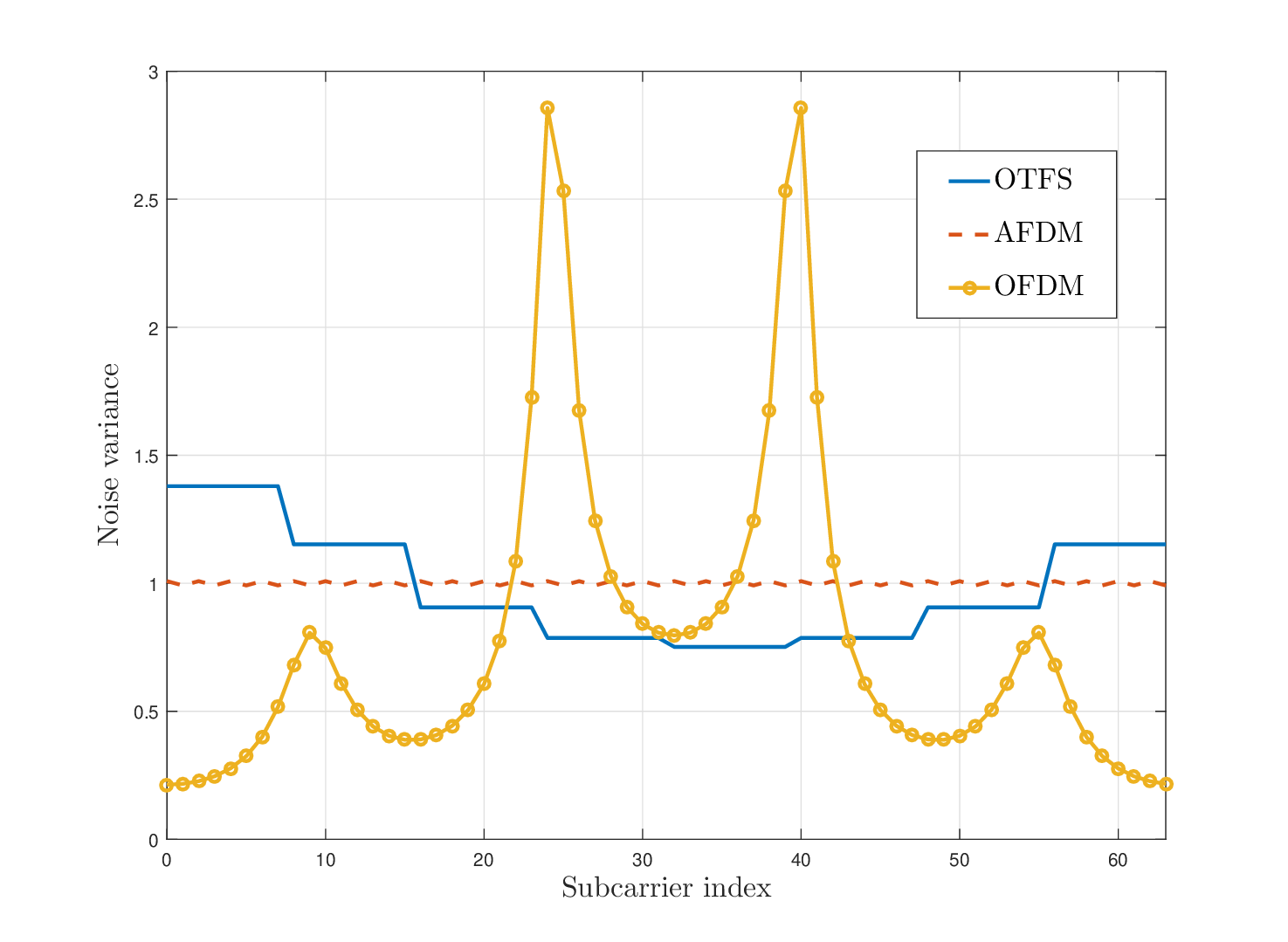}}
\caption{Noise variance versus subcarrier index after demodulation matrix $\textbf{Q}^{-1}$. Comparison of OTFS ($L=8$), AFDM ($q=-4$), and OFDM for: (a) impulse noise, (b) interferer, and (c) equalized noise.}
\label{fig:noisevar}
\end{figure} 

Fig. \ref{fig:noisevar} shows the demodulated noise variance $\mathbb{E}\{|(\textbf{Q}^{-1}\boldsymbol{\Gamma}_f^{1/2} \textbf{w}_w)_m|^2\}$ versus subcarrier index $m=0,1,..,N-1$ for $N=64$ subcarriers for three typical non-white noises: impulsive noise (IN) (a), interferer (b), and equalized noise. The first one (a) has been arbitrarily chosen to show the behavior of the effect of $\textbf{Q}^{-1}$ on an extremly non-white noise. The interferer (b) mimics the presence of distortion on consecutive subcarriers such as an interferer can induce. The equalized noise (c) is the result of a white noise filtered by an equalization matrix. It has been designed with a four-tap channel featuring a constant delay profile. In all three subfigures, the OFDM corresponds to the input non-white noise, since $\textbf{Q}^{-1} = \textbf{I}_N$. Moreover, $K=8$ for OTFS, and $q=-4$ for AFDM. 

It can be observed that the demodulated noise variance of OTFS and AFDM spreads the input noise to other subcarriers. As expected, this is due to the capability of $\textbf{Q}^{-1}$ to whiten the noise. However, to better quantitatively assess this result, Table \ref{tab:tab1} gives the standard deviation $s$ of the demodulated non-white noise variance defined as 

\begin{equation}
s = \left( \frac{1}{N} \sum_{m=0}^{N-1} \left(|(\textbf{Q}^{-1}\boldsymbol{\Gamma}_f^{1/2} \textbf{w}_w)_m|^2 - \bar{m} \right)^2\right)^{1/2}, 
\label{eq:sd}
\end{equation}
where $\bar{m} = \frac{1}{N} \sum_{u=0}^{N-1} |(\textbf{Q}^{-1}\boldsymbol{\Gamma}_f^{1/2} \textbf{w}_w)_m|^2$ is the mean of the variance over the subcarriers. In fact, (\ref{eq:sd}) characterizes the variations of the noise variance samples $|(\textbf{Q}^{-1}\boldsymbol{\Gamma}_f^{1/2} \textbf{w}_w)_m|^2$ around $\bar{m}$. It can be noticed that the standard deviations of OTFS and AFDM are smaller than that of the OFDM (corresponding to the input noise), highlighting the fact that OTFS and AFDM actually whiten the non-white noise. Moreover, $s$ is smaller for AFDM than for OTFS, showing the better ability of AFDM to whiten the noise. This is confirmed by the BER performance later.    

\begin{table}[th]
\centering
\caption{Standard deviation (\ref{eq:sd}) of the variance of the demodulated non-white noise corresponding to Fig. \ref{fig:noisevar}}
\begin{tabular}{|c|c|l|l|}
\hline
\multicolumn{1}{|l|}{$\times 10^{-2}$} & IN   & Interferer & \multicolumn{1}{c|}{\begin{tabular}[c]{@{}c@{}}Equalized\\ noise\end{tabular}} \\ \hline
OFDM                                   & 1.56 & 14.46      & 3.52                                                                           \\ \hline
OTFS                                   & 1.39 & 7.89       & 0.49                                                                           \\ \hline
AFDM                                   & 0.79 & 7.56       & 0.03                                                                           \\ \hline
\end{tabular}\label{tab:tab1}
\end{table}


\subsection{SISO systems}

\begin{figure}[tbp] 
\centering
  \includegraphics[width=\columnwidth]{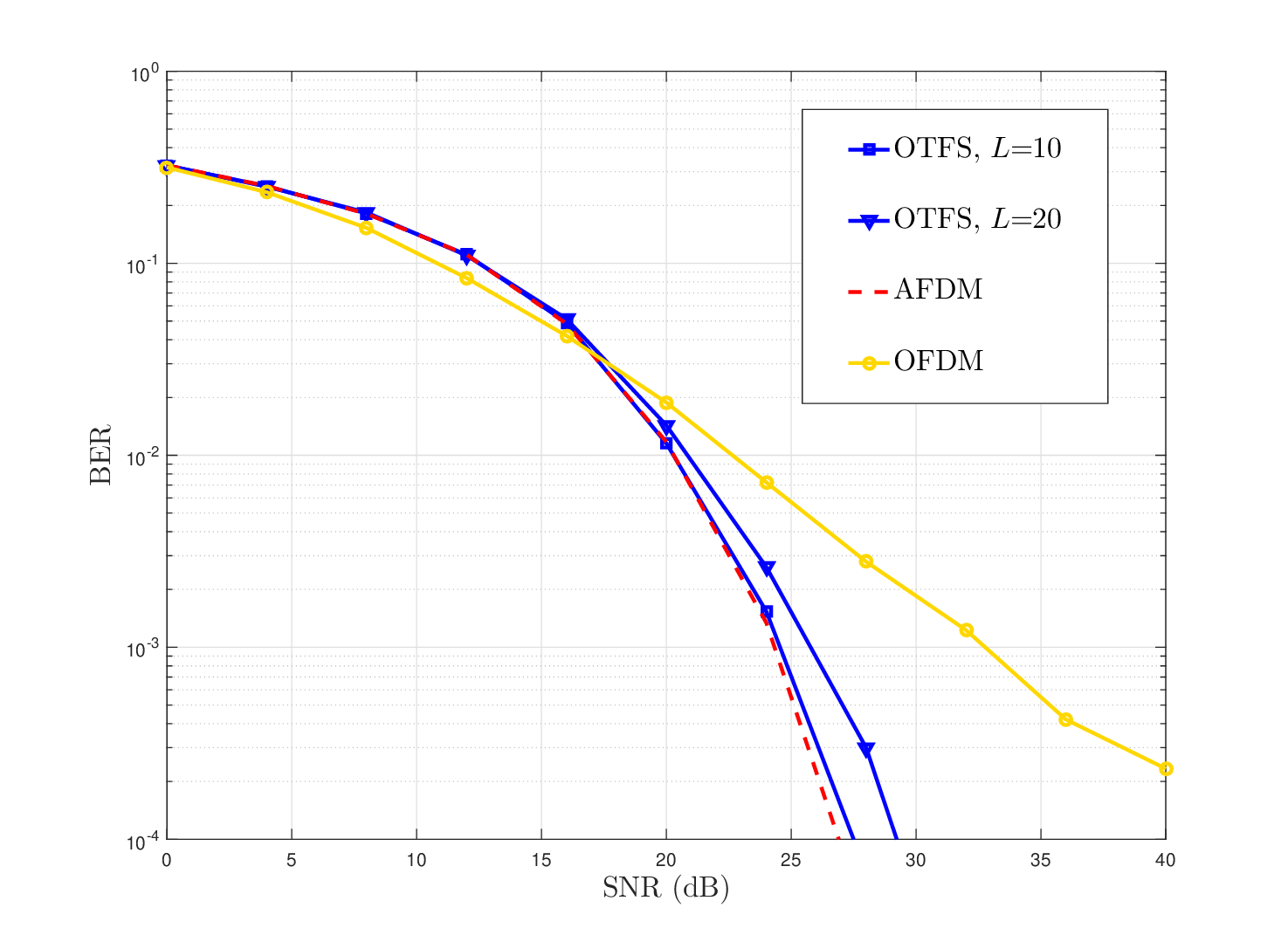} 
\caption{ BER vs SNR (dB) of OTFS ($L=10$ and $L=20$), AFDM, and OFDM in the context of multi-waveform FDMA.  } 
\label{fig:ber_siso} 
\end{figure} 

Fig. \ref{fig:ber_siso} shows the BER versus SNR (dB) of OFDM, OTFS (with $L=10$ and $L=20$), and AFDM ($q=-4$, $\alpha=0.1$) in the context of per-block multi-waveform FDMA. The system consists of $N=1200$ subcarriers divided into $N_1 = N_2 = 600$ subcarriers for each modulation scheme. The subcarrier spacing is set to $30$~kHZ, and the data symbols are randomly taken from a 16-quadrature amplitude modulation (QAM) constellation. The channel used in simulations is composed of $L_c=$8 taps with a uniform delay profile. An MMSE equalizer is used in the frequency domain to mitigate channel distortion. OTFS and AFDM can be seen to outperform OFDM, which is consistent with the literature \cite{bemani21}. However, the performance of OTFS decreases when $L$ increases (a loss of 1 dB can be observed between $L=10$ and $L=20$), such as supposed in the sparsity analysis of $\textbf{Q}^{-1}$ in (\ref{eq:QOTFS1})-(\ref{eq:otfsnoise}).

\begin{figure}[tbp]
\centering
  \subfigure[BER versus $L$ (log).]{\includegraphics[width=\columnwidth]{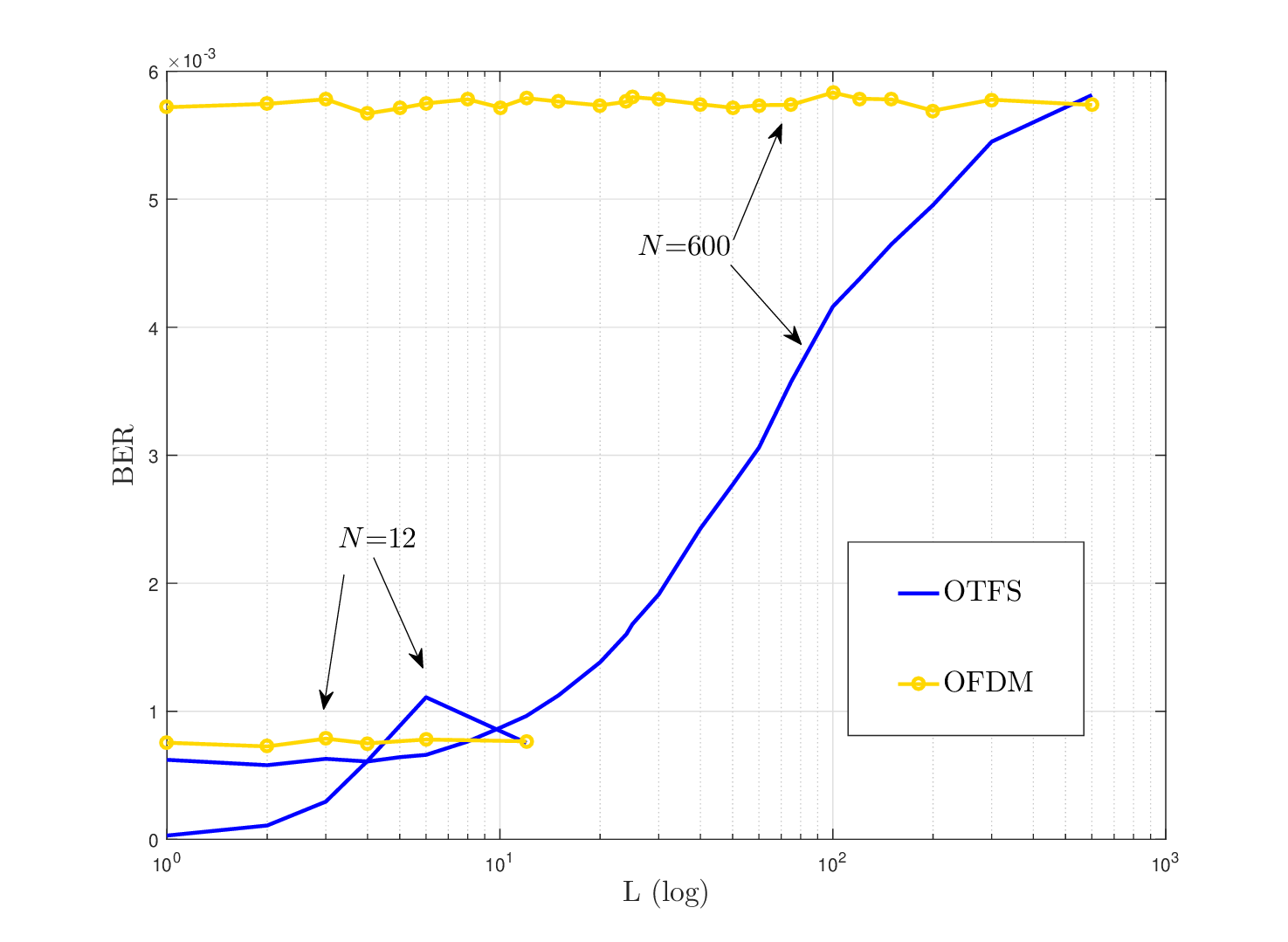}} \quad
  \subfigure[BER versus $q$.]{\includegraphics[width=\columnwidth]{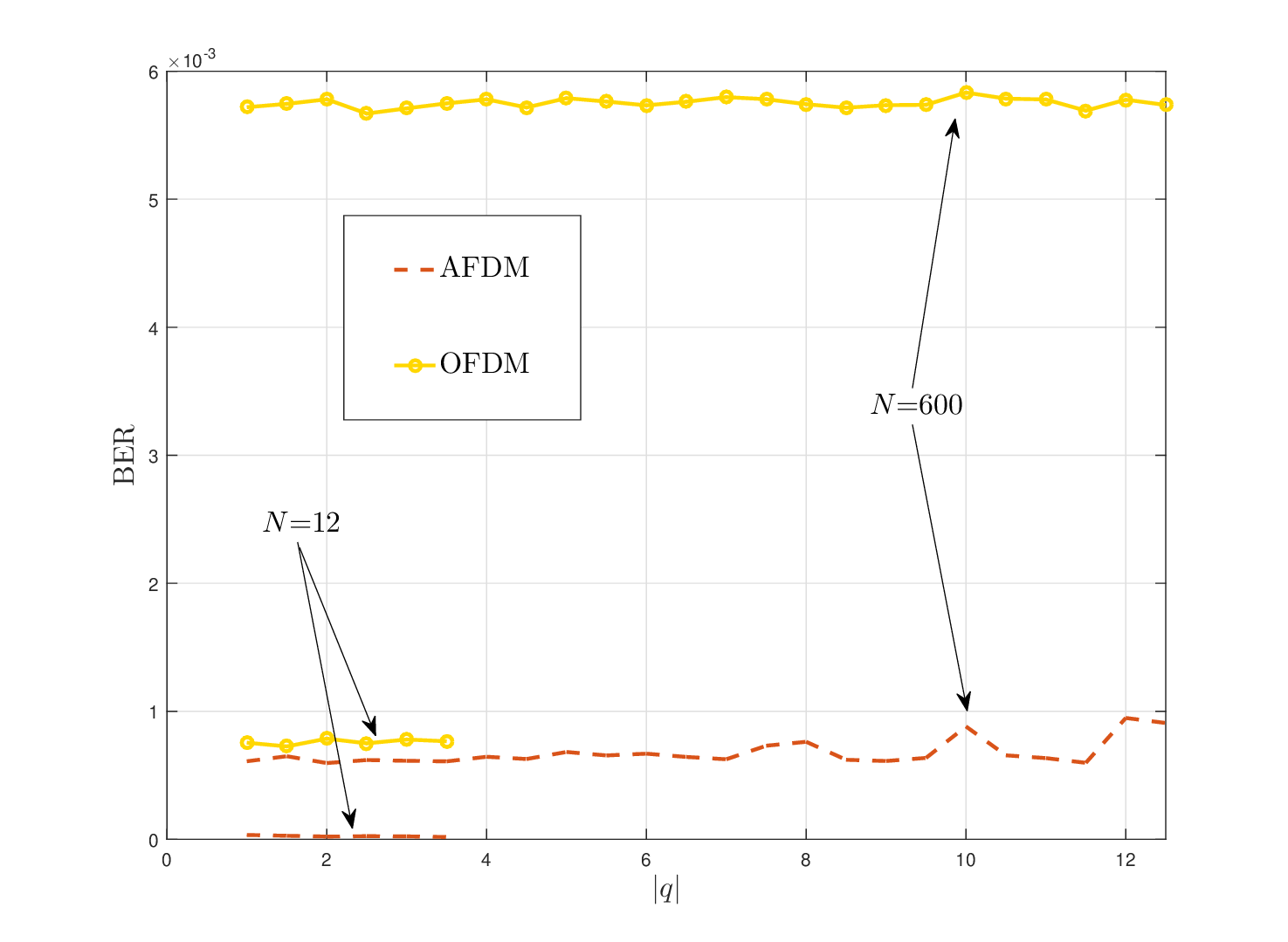}}
\caption{BER performance versus L (a) and $q$ (b) for SNR=25dB. Comparison of OFDM and OTFS (a), and OFDM and AFDM (b). }
\label{fig:compareparamLq}
\end{figure}

To confirm this result, the BER performance of OTFS and AFDM is tested against $L$ (a) and $q$ (b), respectively, and compared with OFDM as a reference, for a SNR of 25dB in Fig. \ref{fig:compareparamLq}. Two values of $N \in \{12,600\}$ are considered, corresponding to narrowband and broadband signals, respectively. The rest of the parameters remains the same as previously. According to the analysis of the sparsity of $\textbf{Q}^{-1}$, the BER of OTFS increases with $L$ to reach that of OFDM when $L=N$, and it is minimum for low values of $L$. In particular, for $N=12$, the minimum is achieved only for $L=1$, which is limiting for narrowband applications. At the same time, the BER of AFDM remains almost constant for any considered value of $q$, and corresponds to the minimum of OTFS. This shows the superiority of AFDM against OTFS in frequency selective channel for any chosen parameter (the same can be observed for other non-white noise models), and is also consistent with the diversity analysis \cite{rou24}: full diversity depends on conditions on $L$ and $K$ (and then $N$) in OTFS to cope with delay and Doppler, where it only depends on $q$ in AFDM to cope with Doppler shift, independently of $N$.

\begin{figure}[tbp] 
\centering
  \includegraphics[width=\columnwidth]{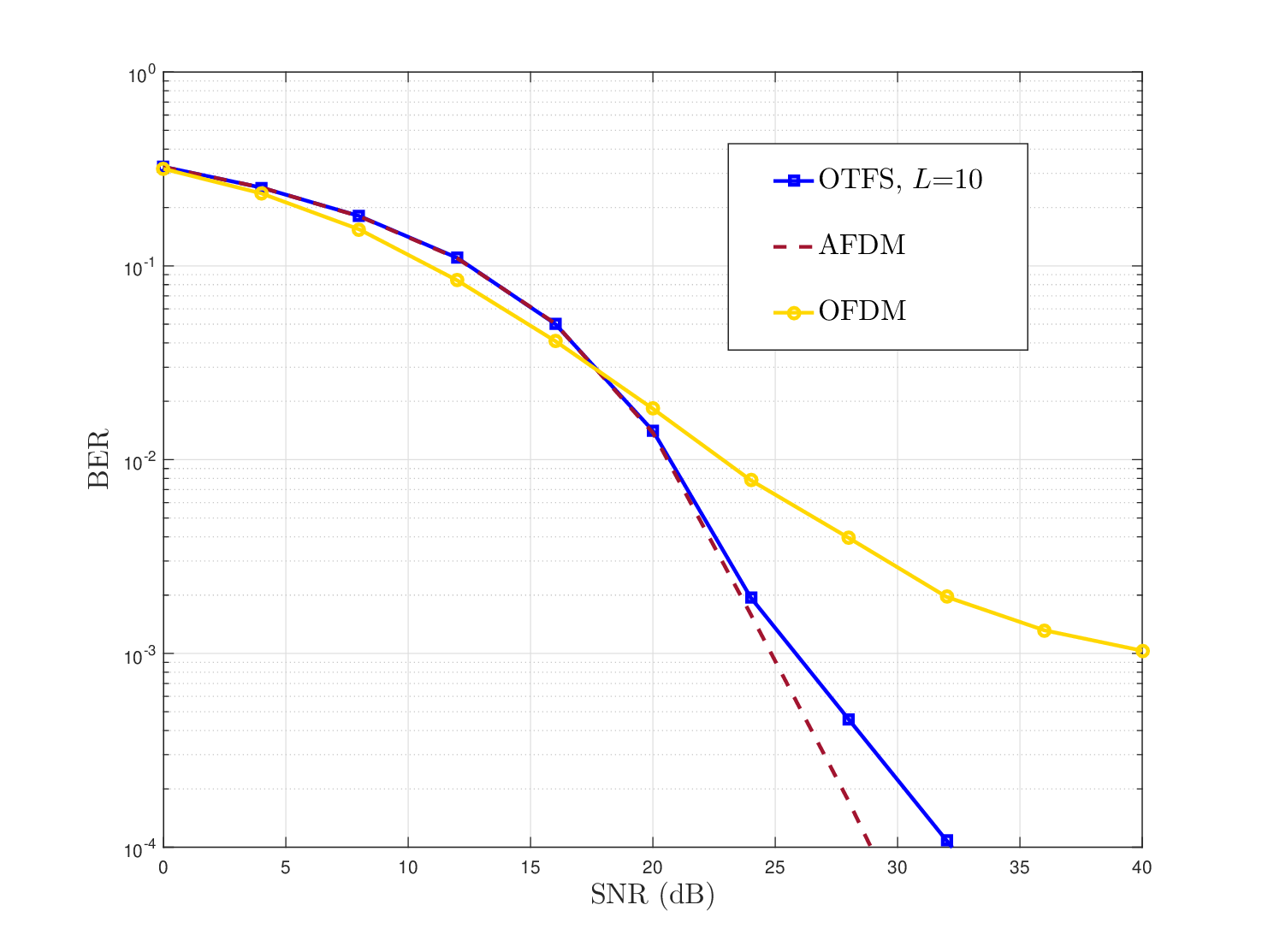} 
\caption{ BER vs SNR (dB) of OTFS ($L=10$ and $L=20$), AFDM, and OFDM in a doubly dispersive channel with impulse noise.  } 
\label{fig:chan_imp_noise} 
\end{figure}

In Fig. \ref{fig:chan_imp_noise}, we investigate to what extent the approximation of diagonal $\boldsymbol{\Gamma}_f$ holds in the whitening analysis proposed in Section \ref{sec:perf_analysis}. To this end, BER versus SNR (dB) using the same parameters as in Fig. \ref{fig:ber_siso} is considered, except that a doubly dispersive channel with maximum Doppler shift $\theta_l=0.3$ is used. Moreover, an impulsive noise as in Fig. \ref{fig:noisevar}-(a) is also applied. The number of subcarriers is set to $N=600$, and the modulation parameters are $L=10$ for OTFS and $q=-4$ for AFDM. In Fig. \ref{fig:ber_siso}, it has been observed that the performance of OTFS ($L=10$) and AFDM match. Fig. \ref{fig:chan_imp_noise} shows that AFDM outperforms OTFS in the highest SNR range. This behavior can be explained by the fact that the AFDM demodulation better whiten the non-white noise (simulated by the impulse noise) compared with OTFS, even in presence of non-diagonal $\boldsymbol{\Gamma}_f$ matrix (simulated by the doubly dispersive channel). This result clearly validates the proposed performance analysis based on noise whitening.   

\section{Conclusion}
\label{sec:conclusion} 

This paper presents a performance analysis of AFDM and OTFS against non-white noise. The latter is used to model different sources of disturbance, such as the equalization of frequency selective channel, filtering of white noise at the receiver, aliasing, interference, or jamming. It is shown that the robustness of the waveforms against non-white noise can be directly linked to the capability of their demodulation matrix to whiten the additive noise, which is in turn related to the sparsity of these matrices. The analysis reveals that AFDM should outperform OTFS because its demodulation matrix is less sparse than that of OTFS. Based on this, different applications and use cases involving AFDM and OTFS are presented, including coexistence with OFDM signal. Finally, the theoretical results applied in the presented use cases are verified through simulation results, which show the superiority of AFDM compared to OTFS and OFDM against non-white noise.    



\appendix

\section{Proof of Non-Sparse Matrix $\textbf{Q}^{-1}$ in AFDM}
\label{app:appendixA}

The (non-)sparsity of $\textbf{Q}^{-1}$ is equivalent to that of $\mathbfcal{F}_{N}\boldsymbol{\Lambda}_q^H \mathbfcal{F}_{N}^H$ in (\ref{eq:QAFDM2}). We hereby show that $\mathbfcal{F}_{N}\boldsymbol{\Lambda}_q^H \mathbfcal{F}_{N}^H$ is generally non-sparse. To this end, we first limit to $q \in \mathbb{Q}$ instead of $q \in \mathbb{R}$, which is relevant in digital systems with inherent limited precision. Then, $q$ can be approximated with any expected precision as $q = \frac{a}{b}$, where $a$ and $b$ are integers. The generalized quadratic Gauss sum in (\ref{eq:QAFDM2}) then yields 

\begin{equation}
(\mathbfcal{F}_{N}\boldsymbol{\Lambda}_q^H \mathbfcal{F}_{N}^H)_{u,0} = \frac{1}{\sqrt{N}} \sum_{k=0}^{N-1} e^{-j\pi \frac{ak^2}{bN}} e^{-2j\pi \frac{ku}{N}}. 
\label{eq:QAFDM3}
\end{equation} 
Assessing the DFT in (\ref{eq:QAFDM3}) is equivalent to performing the DFT of size $bN$ of the discrete function 

\begin{equation}
e^{-j\pi \frac{ak^2}{bN}} \Pi_{[0,N-1],k}, 
\end{equation}
followed by a decimation of factor $b$, with $k=0,1,..,bN-1$, and $\Pi_{[0,N-1],k}$ is the rectangular function such that $\Pi_{[0,N-1],k} = 1$ if $0\leq k\leq N-1$, and $\Pi_{[0,N-1],k} = 0$ else. Hence, $(\mathbfcal{F}_{N}\boldsymbol{\Lambda}_q^H \mathbfcal{F}_{N}^H)_{u,0}$ can be rewritten as 

\begin{align}
    (\mathbfcal{F}_{N}\boldsymbol{\Lambda}_q^H \mathbfcal{F}_{N}^H)_{u,0} =& (\textbf{D}_{b}\mathbfcal{F}_{bN}\boldsymbol{\Lambda}_a^H \boldsymbol{\Pi}_{[0,N-1]} \mathbfcal{F}_{bN}^H)_{u,0} \nonumber \\
    =& (\textbf{D}_{b}\mathbfcal{F}_{bN}\boldsymbol{\Lambda}_a^H\mathbfcal{F}_{bN}^H \mathbfcal{F}_{bN}\boldsymbol{\Pi}_{[0,N-1]} \mathbfcal{F}_{bN}^H)_{u,0}, 
    \label{eq:QAFDM4}
\end{align}
where $\textbf{D}_{b}$ is the $N \times bN$ decimation matrix, and $\boldsymbol{\Lambda}_a^H$ and $\boldsymbol{\Pi}_{[0,N-1]}$ are the $bN \times bN$ diagonal matrices that contain the samples $e^{-j\pi \frac{ak^2}{bN}}$ and $\Pi_{[0,N-1],k}$, respectively. It is worth noticing that the matrix multiplication (except the decimation) in (\ref{eq:QAFDM4}) corresponds to the convolution between two circulant matrices $\mathbfcal{F}_{bN}\boldsymbol{\Lambda}_a^H\mathbfcal{F}_{bN}^H$ and $\mathbfcal{F}_{bN}\boldsymbol{\Pi}_{[0,N-1]} \mathbfcal{F}_{bN}^H$. The elements of the first column of $\mathbfcal{F}_{bN}\boldsymbol{\Lambda}_a^H\mathbfcal{F}_{bN}^H$ is given, for any $u=0,1,..,bN-1$, by 

\begin{equation}
(\mathbfcal{F}_{bN}\boldsymbol{\Lambda}_a^H\mathbfcal{F}_{bN}^H)_{u,0} = \frac{1}{\sqrt{bN}} \sum_{k=0}^{bN-1} e^{-j\pi \frac{ak^2}{bN}} e^{-2j\pi \frac{ku}{bN}}.  
\label{eq:QAFDM5}
\end{equation}
Note that (\ref{eq:QAFDM5}) simplifies only for certain value of $a$, $b$, and $N$, such as presented in \cite{savaux24ieeetcom}. In general, we can conclude that $\mathbfcal{F}_{bN}\boldsymbol{\Lambda}_a^H\mathbfcal{F}_{bN}^H$ is not sparse. In turn, the elements of the first column of $\mathbfcal{F}_{bN}\boldsymbol{\Pi}_{[0,N-1]} \mathbfcal{F}_{bN}^H$ is given, for any $u=0,1,..,bN-1$, by 

\begin{equation}
(\mathbfcal{F}_{bN}\boldsymbol{\Pi}_{[0,N-1]} \mathbfcal{F}_{bN}^H)_{u,0} = \frac{e^{-j\pi u (\frac{1}{b}-\frac{1}{bN})}}{\sqrt{bN}}\times \frac{\sin\left( \pi \frac{u}{b}\right) }{\sin\left( \pi \frac{u}{bN}\right)}.
\end{equation}
Once again, $\mathbfcal{F}_{bN}\boldsymbol{\Pi}_{[0,N-1]} \mathbfcal{F}_{bN}^H$ is not sparse. It results that in general, $\mathbfcal{F}_{N}\boldsymbol{\Lambda}_q^H \mathbfcal{F}_{N}^H$ and in turn $\textbf{Q}^{-1}$ is not sparse and should then be able to whiten the received non-white noise. 

\bibliographystyle{IEEEtran}
\bibliography{mabiblio}
\end{document}